\documentclass[9pt,twocolumn,twoside]{pnas-new}
\templatetype{pnasresearcharticle} 

\usepackage{graphicx}
\usepackage{natbib}
\usepackage[section] {placeins}
\usepackage{subfigure}
\usepackage{float}
\usepackage{color}
\usepackage{amsmath}

\usepackage[scaled]{helvet}

\usepackage[T1]{fontenc}
\usepackage{color}

\def\apj{ApJ}
\def\apjl{ApJL}                
\def\aap{A\&A}                
\def\aaps{{A\&AS}}              
\def\mnras{{MNRAS}}             
\def\nat{{Nature}}              

\def\msun{{\rm\,M_\odot}}

\newcommand{\mpc}{\, {\rm Mpc}}

\def\h2{${\rm\,H_2}$}

\newcommand{\cm}{\mathrm{cm}}

\newcommand{\alf}{Alfv$\acute{\text{e}}$n\ } 

\makeatletter 

\def\mpc{{\rm\,Mpc}}

\def\msun{{\rm\,M_\odot}}

\def\vol#1  {{{#1}{\rm,}\ }}

\newcount\refno
\newcount\rfno
\def\eq{$^{\the\refno\ }$\advance\refno by 1}
\def\ad{\advance\rfno by 1}

\def\clock{\count0=\time \divide\count0 by 60
     \count1=\count0 \multiply\count1 by -60 \advance\count1 by \time
     \number\count0:\ifnum\count1<10{0\number\count1}\else\number\count1\fi}

\def\myputfigure#1#2#3#4#5%
{\hskip0.03\textwidth\vskip#5pt
\makebox[0pt]{\hskip#2in
\includegraphics[width=#3\textwidth]{#1}}\vskip#4pt\hfill}

\def\newblock{\hskip .11em plus .33em minus .07em}

\begin{document}

\title{Global Preventive Feedback of Powerful Radio Jets on Galaxy Formation}

\author[a,1,2]{Renyue Cen}

\affil[a]{Center for Cosmology and Computational Astrophysics, Institute for Advanced Study in Physics,
and Institute of Astronomy, School of Physics, Zhejiang University, Hangzhou 310027, China; renyuecen@zju.edu.cn}

\leadauthor{Cen}

\significancestatement{
Negative feedback processes from the growth of supermassive black holes are believed to play a central role to the formation of galaxies.
Internal feedback processes by essentially driving gas away from galaxies
have been widely used in cosmological simulations. 
In this article, we investigate a new type of negative feedback process,
termed external negative feedback,
due to powerful radio jets by energizing intergalactic medium.
The intergalactic medium endowed with significant magnetic fields 
is retarded or prevented from entering subsequent halos.
This external feedback process is in a sense pro-active and preventative, 
in contrast to the case of interval feedback processes.
Inclusion of this external feedback process in
the next generation of cosmological simulations may be imperative.
}

\authorcontributions{Please provide details of author contributions here.}
\authordeclaration{The authors declare no conflict of interest.}
\equalauthors{\textsuperscript{1}Author contributions: R.C. designed, performed 
research and wrote the paper.}
\correspondingauthor{\textsuperscript{2}To whom correspondence should be addressed. E-mail: renyuecen@zju.edu.cn}


\begin{abstract}
Firmly anchored on observational data, giant radio lobes from massive galaxies hosting supermassive black holes can exert a major negative feedback effect,
by endowing the intergalactic gas with significant magnetic pressure hence retarding or preventing gas accretion onto less massive halos in the vicinity.
Since massive galaxies that are largely responsible for producing the giant radio lobes, this effect is expected to be
stronger in more overdense large-scale environments, such as proto-clusters, than in underdense regions, such as voids. 
We show that by redshift $z=2$ halos with masses up to 
$(10^{11-12}, 10^{12-13})\msun$ 
are significantly hindered from accreting gas due to this effect
for radio bubble volume filling fraction of $(1.0, 0.2)$, respectively.
Since the vast majority of the stars in the universe at $z<2-3$ form precisely in those halos, this negative feedback process is likely one major culprit for 
causing the global downturn in star formation 
in the universe since. It also provides a natural explanation 
for the rather sudden flattening of the slope of
the galaxy rest-frame UV luminosity function around $z\sim 2$.
A cross-correlation between proto-clusters and Faraday rotation measures
may test the predicted magnetic field. 
Inclusion of this external feedback process in
the next generation of cosmological simulations may be imperative.
\end{abstract}

\dates{This manuscript was compiled on \today}
\doi{\url{www.pnas.org/cgi/doi/10.1073/pnas.XXXXXXXXXX}}

\maketitle
\thispagestyle{firststyle}
\ifthenelse{\boolean{shortarticle}}{\ifthenelse{\boolean{singlecolumn}}{\abscontentformatted}{\abscontent}}{}

\firstpage{2}

\dropcap{C}osmological hydrodynamic simulations have entered its fourth decade since the pioneering works in the
late eighties and early nineties of the last century
\citep[e.g.,][]{1989Chiang,1990Cen,1990Evrard,1991Katz}.
In the first two decades, the focus was largely on the evolution of the intergalactic medium, regions significantly removed from star formation in galaxies,
with a number of notable successes, including the finding
of the fluctuating nonlinear Gunn-Peterson cosmological density field
as the origin of the Lyman alpha forest
\citep[][]{1994Cen, 1995Zhang, 1996Hernquist, 1996MiraldaEscude},
a successful account of the missing baryons in the universe at zero redshift \citep[][]{1999Cen, 2001Dave} and the discovery of two modes of gas accretion into galaxies
\citep[][]{2005Keres}.
In the most recent two decades, 
driven in part by the availability of large computing power, 
we witnessed large-scale cosmological simulations
with increasingly high numerical resolutions and ever more sophisticated implementations
for feedback processes from stellar evolution and growth of supermassive black holes
\citep[e.g.,][]{2014Vogelsberger,2014Genel,2015Schaye,2015Crain,2015Schaller,2015Trayford,2016McAlpine,
2018Pillepich,2018Springel,2018Nelson, 2019Dave,2022Sorini}.
On the feedback processes from AGN, the mainstream implementations of various modes 
(QSO or radio mode) can be classified as internal feedback processes by pushing entered gas away from galaxies.

In the {\it article} we put forth an important AGN energy source,
namely, the powerful FR II radio jets,
for regulating the thermodynamic state of the cosmic gas, especially in the low-density intergalactic medium (IGM).
FR II jets are observed to transport a large amount of energy to megaparsec scale, deep into the low-density IGM
\citep[e.g.,][]{2008Machalski}.
Preferentially affecting low-density gas is economical energetically by maximizing entropy generation.
In contrast to internal feedback processes currently employed in cosmological simulations,
this new feedback process is of global, collective and multi-generational nature, 
where radio lobes generated by earlier galaxies will
retard, reduce or prevent subsequent gas accretion onto galaxies in the neighborhood.

\section*{Suppression of Galactic Gas Accretion due to Intergalactic Magnetic Field}

We consider the simplest case that
powerful AGN radio jets send magnetic fields permeating the entire universe
but will discuss later if filling fraction is less than unity, the likely case.
Let us denote the mean magnetic field of the universe due to AGN jets as $B_i$ at certain redshift.
Then under the adiabatic assumption,
when a small gas parcel is eventually brought to the virial surface of a halo with a density
equal to the virial overdensity $\delta_v$,
the magnetic field would have increased to $B_v$,
\begin{equation}
\label{eq:Bv}
B_v = B_i \delta_v^{2/3}.
\end{equation}
\noindent
Can this gas parcel enter this virial radius?
To answer this question, let us make a most conservative assumption that this gas
has completely lost its thermal pressure when arriving at the virial surface of the halo.
Under a likely valid assumption that it is not self-gravitating
(because otherwise it would have entered a halo already),
the magnetic pressure force is dynamically analagous to thermal pressure force
in terms of retarding the gas from accreting onto the halo.
When the magnetic pressure force exceeds gravitational force due to the dark matter halo
at the virial radius, or equivalently, the magnetic temperature,
defined as $T_B\equiv B_v^2/8\pi n_v k_B$, where $n_v$ is gas number density at the virial radius and $k_B$ the Boltzmann's constant,
exceeds the virial temperature of the halo,
the gas parcel will be prevented from entering the halo.
This statement can be expressed in terms of a threshold \alf velocity $v_A=\sqrt{B_v^2/4\pi\rho_v}$ (where $\rho_v=m_pn_v$)
\begin{equation}
\label{eq:Bv2}
\begin{split}
v_A = \sqrt{2} \sigma_v,
\end{split}
\end{equation}
\noindent
where $\sigma_v$ is the 1-d velocity dispersion of the halo.

\begin{figure}
\centering
\includegraphics[width=1.0\linewidth]{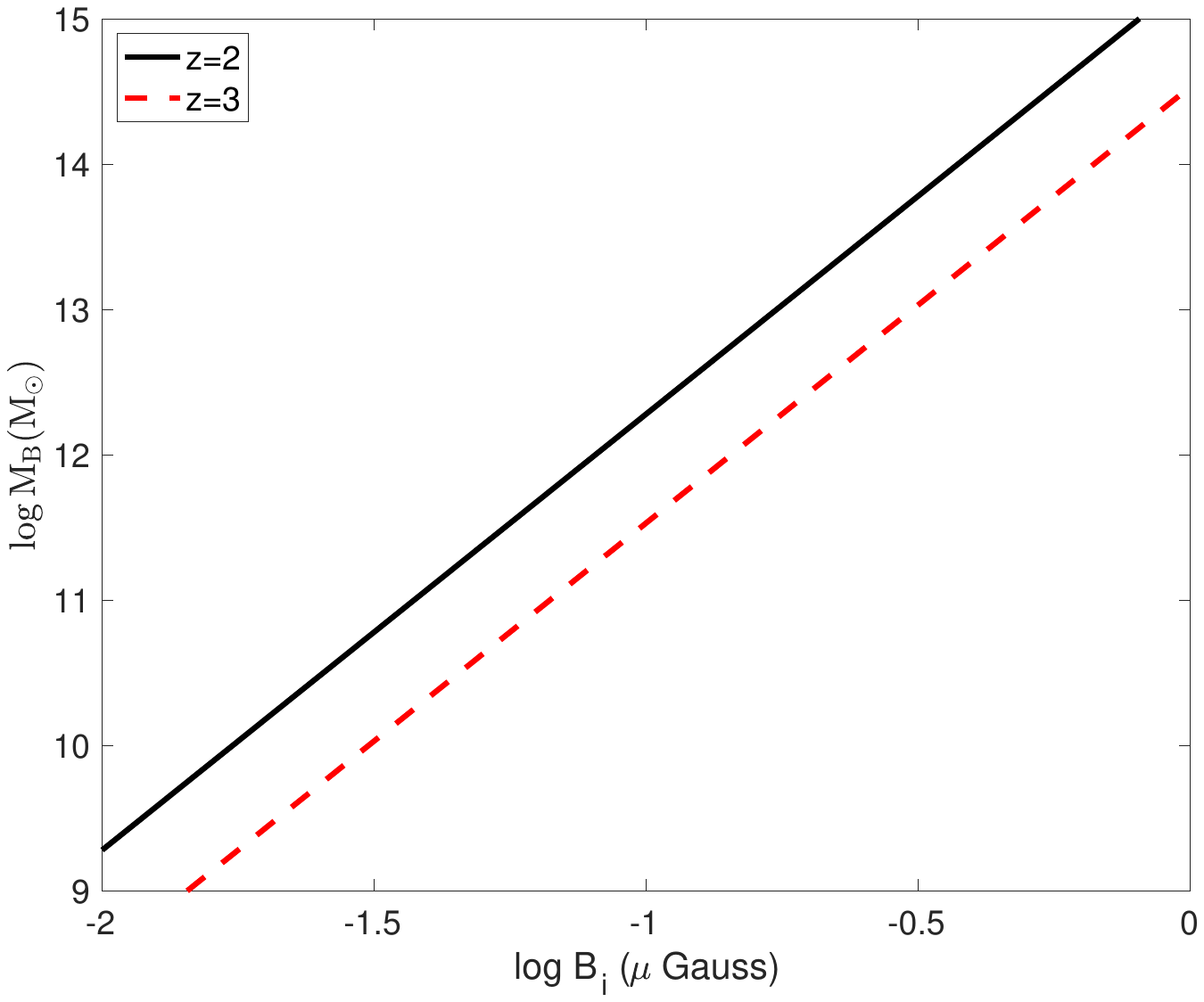}
\caption{shows the magnetic Jeans mass, $M_B$, as a function of the magnetic field in the mean density of the universe, $B_i$,
at $z=2$ (black solid curve) and $z=3$ (red dashed curve).}
\label{fig:MhB}
\end{figure}

Combining Eq (\ref{eq:Bv}) and (\ref{eq:Bv2}), along with the relation between halo mass and $\sigma_v$,
we obtain the results shown in Figure (\ref{fig:MhB}).
A more precise formulation may be quantified with detailed simulations, in a way similar to the
Jeans filtering effect demonstrated insightfully in \cite[][]{2000Gnedin} and is reserved
for a future study.
Nevertheless, we see that, if an initial mean field $B_i$ of order $0.1\mu$G is reached,
gas accretion onto halos as massive as $M_h\sim 10^{12}\msun$ will be significantly affected.
In the next section, we examine whether such a field may be injected into the intergalactic medium
by powerful radio jets.

\section*{Intergalactic Magnetic Field Due to Powerful Radio Jets}

To estimate the energetics of powerful radio jets, we introduce a few variables.
We denote the global ratio of supermassive black hole (SMBH) mass to
stellar mass as $\beta$ adopting $0.002$ measured locally \citep[e.g.,][]{2003Marconi},
although there is preliminary evidence that this ratio may increase with increasing redshift \citep[e.g.,][]{2020Ding}.
We denote the fraction of SMBH rest mass energy ($M_{\rm SMBH} c^2$) released in the form of  powerful radio jets 
as $\eta_J$ for the radio loud galaxy population,
and $\eta_R$ as the fraction of all relevant galaxies classified as radio loud galaxies.
To obtain the fraction of jet energy in the form of magnetic energy,  $\eta_B$,
we have performed MHD simulations of magnetic energy powered 
explosion and find that for a point injection of $10^{61}$erg magnetic energy,  $\eta_B$ retains a value of $15\%(0.5Mpc/R_B)$ at late stages,
where $R_B$ is the bubble radius.
The universal stellar mass density formed by a redshift of interest in units of the total baryonic density
is denoted as $\eta_*$.
With these variables and assuming that the radio jets send magnetic fields to large distances to fill a volume fraction 
of the universe, $f$,
we obtain the mean magnetic field energy density
\begin{equation}
\label{eq:Bi}
\begin{split}
{B_i^2\over 8\pi} &= 3\times 10^{-4}\left({\eta_{\rm M}\over 0.002}\right) \left({\eta_B\over 0.15}\right)\left({\eta_R\over 0.2}\right)\left({f\over 0.2}\right)^{-1}\eta_* \eta_J c^2 \rho_b(z) \\
&=2.8\times 10^{-15}(\mu G)^2\left({\eta_{\rm M}\over 0.002}\right) \left({\eta_B\over 0.15}\right)\left({\eta_R\over 0.2}\right)\left({f\over 0.2}\right)^{-1} \\
&\quad\quad\quad\quad\quad\quad\quad\quad\quad\quad \left({\eta_*\over 0.018}\right) \left({\eta_J\over 0.05}\right)\left({1+z\over 3}\right)^3,
\end{split}
\end{equation}
\noindent
where $c$ is the speed of light, $\rho_b(z)$ the mean baryonic density at the redshift in question.

\begin{figure}
\centering
\includegraphics[width=1.0\linewidth]{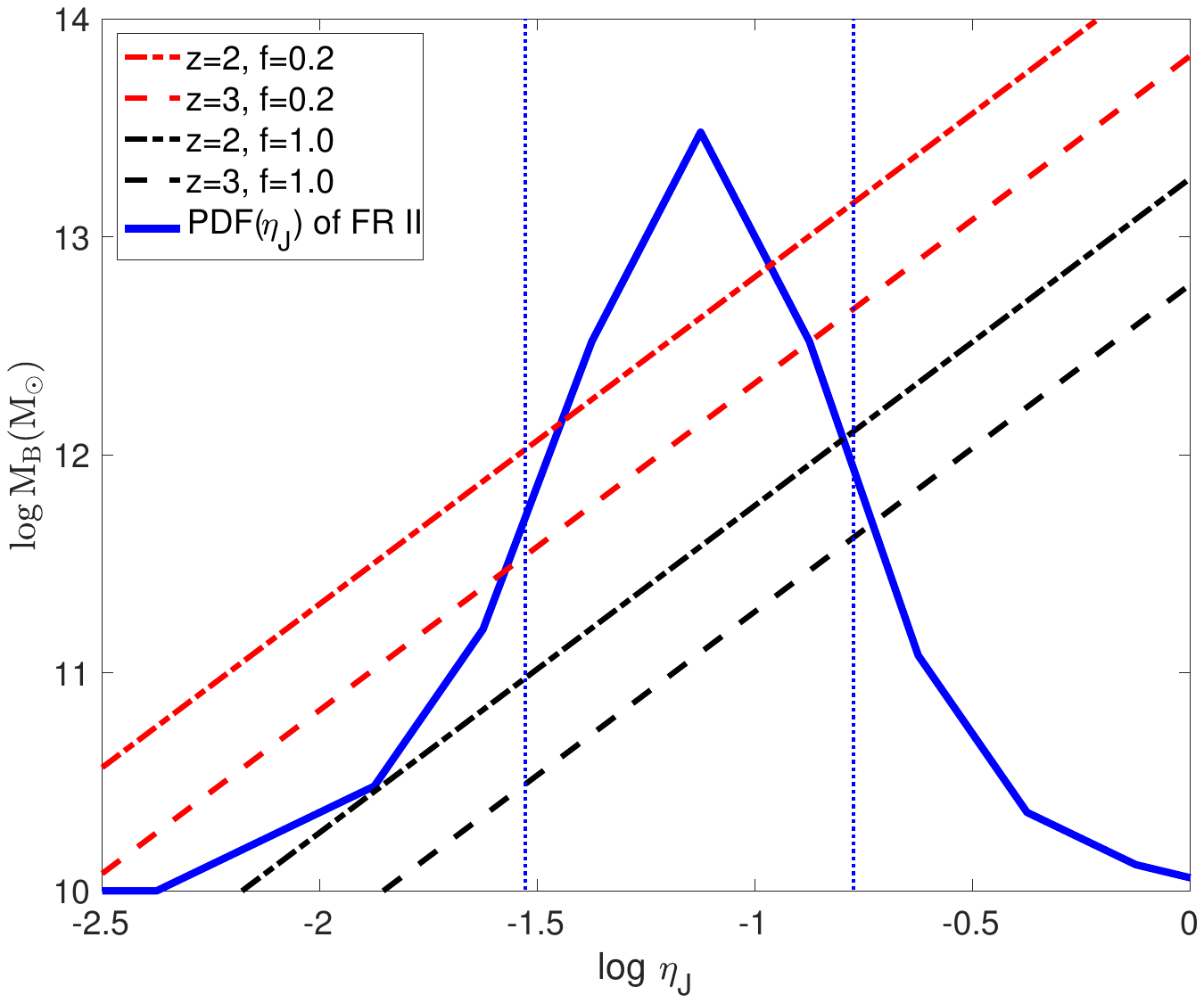}
\caption{shows the magnetic Jeans mass, $M_B$, as a function of
$\eta_J$, the fraction of the SMBH rest mass energy released in the form of powerful radio jets
in radio-loud galaxies, 
at $z=2$ (dot-dashed lines) and $z=3$ (dashed lines).
Two cases of filling factor $f$ are shown:
the two red lines are for $f=0.2$, 
whereas the two black lines are for $f=1.0$.
Other relevant parameters are as follows.
The observed mean stellar mass density of $5.7\times 10^7\msun~{\rm Mpc}^{-3}$ and
$3.6\times 10^7\msun~{\rm Mpc}^{-3}$ at $z=2$ and $z=3$, respectively, are from
\cite[][]{2023Weaver}.
For the fraction of all relevant galaxies classified as radio loud galaxies, $\eta_R=0.2$ is adopted \citep[][]{2023Best}. 
The fraction of original jet magnetic energy remaining in the form of magnetic energy 
in the final giant radio bubbles is assumed to be $\eta_B=7.5\%$, estimated from simulations.
Also shown as solid blue curve is the observed probability density distributions of $\eta_J$ for 
 the classic FR II sources \citep[][]{2021Daly}.}
\label{fig:Mhfjet}
\end{figure}

Let us now go through the numerical values of various variables, based solely on observational data.
We adopt the universal stellar density of 
($5.7\times 10^7\msun~{\rm Mpc}^{-3}$,  
$3.6\times 10^7\msun~{\rm Mpc}^{-3}$) (in comoving volume) at $z=(2,3)$, respectively,
from \citep[][]{2023Weaver} converted to values corresponding to the Salpeter initial mass function,
yielding $\eta_*=(0.018,0.011)$.
From Figure 10 of \citep[][]{2023Best} we find radio loud fraction
$\eta_R=(0.19,0.24)$ at $z=(3,2)$, respectively;
given the uncertainties involved, we shall adopt a single value of $\eta_R=0.2$,
which is a good approximation for the entire redshift range of $z=0-4$.
The distribution of $\eta_J$ for the classic FR II sources (shown as blue solid curve)
is from \citep[][]{2021Daly}.
If $f=1$, we find $B_i=0.12\mu$G at $z=2$, which is the mean magnetic field strength over the entire
universe.

We express the threshold halo mass $M_B$ due to magnetic pressure,
as a function of $\eta_J$, shown in Figure (\ref{fig:Mhfjet}), where $\eta_B=0.075$ is adopted.
The observed distribution of $\eta_J$ for the FR II sources
shown as the blue solid curve has the 
$\pm 1\sigma$ range indicated by the two vertical dotted blue lines.
$M_B$ values at the two crossing points between an inclined line and
the two vertical dotted blue lines indicate the $\pm 1\sigma$ range 
of $M_B$ for that line.
We find $\pm 1\sigma$ range of 
$[2.2\times 10^{10},4.0\times 10^{11}]\msun$ and
$[1.0\times 10^{11},1.3\times 10^{12}]\msun$ 
at $z=3$ and $z=2$, respectively, assuming $f=1.0$;
$\pm 1\sigma$ range becomes
$[3.5\times 10^{11},4.5\times 10^{12}]\msun$ and
$[1.1\times 10^{12},1.4\times 10^{13}]\msun$ 
at $z=3$ and $z=2$, respectively, if $f=0.2$.

\section*{Discussion and Predictions}

We have shown that, with a set of parameters anchored firmly on observational data,
the magnetic pressure of accreting gas infused with magnetic fields originating from
giant radio lobes exerts significant retarding effects on gas accretion into halos
as massive as $10^{12-12.5}\msun$.
Here we discuss possible other effects and tests of this physical process.

\subsection*{Effect of Parker Instability}

For a dark matter halo the density profile near the virial radius $r_v$ has a local slope $\alpha\sim -2.3$.  Then, it can be shown that the minimum wavelength of the planar undular instability at the virial radius $r_v$
is $\lambda_{\rm min} = 4 \sqrt{2/(-1-\alpha)} (v_A/\sqrt{2}\sigma_v) r_v = 5.0 (v_A/\sqrt{2}\sigma_v) r_v$. Since $v_A/\sqrt{2}\sigma_v$ (see Eq 2) is about unity or larger for the magnetic suppression to be significant, we see that the perturbation with the minimal wavelength, which is also the fastest growing mode, grows on a time scale of order five times the dynamic time of the halo at the virial radius.  Hence, the induced RT instability, if it exists, will unlikely be very important, since the relevant time scale close a halo is its dynamical time. In other words, if the gas eventually becomes unstable, the long waiting time of five dynamical times is already a significant retarding effect. Moreover, $\lambda_{\rm min}$ is much larger than $r_v$ and in fact larger than one half of the circumference, such a planar undular instability then would not exist in the first place.  Thus, we expect the RT instability due to magnetic buoyancy is unlikely to significantly alter the proposed magnetic pressure based gas accretion suppression onto eligible halos.

\subsection*{Magnetic Reconnection}

In our model, the initial magnetic field that we focus on is on large scales, of order 1Mpc, larger than affected halos of virial radius of 300kpc or smaller. 
Hence the magnetic field reaching a halo is on the scale of the halo size or larger. Given the normalized reconnection rate of order 0.1, inferred from observations and from theoretical calculations, it implies that the magnetic reconnection time is about ten times the system time (i.e., halo dynamical time), which is on the order of Hubble time at any redshift.  Moreover, when the gas is en route to the halo, the reconnection time scale is still longer. This indicates that magnetic reconnection, which may take place to perhaps cancel some of the smaller scale stressed magnetic field on relatively shorter time scales, is unlikely to reduce significantly the overall magnetic energetics of gas endowed with large-scale magnetic field with $v_A = \sqrt{2}\sigma$ or larger.

\subsection*{Volume Filling Fraction of Radio Bubbles}

We now return to the issue of volume filling fraction of the magnetic bubbles, $f$.
We can only give a very rough estimate given uncertainties.
Let us assume that the radio AGNs are dominated by massive galaxies, based on available observational evidence \citep[e.g.,][]{2021Daly}.
To facilitate an estimation, let us assume that $0.5L_*$ galaxies and above are responsible for making the giant radio lobes;
the mean separation of $0.5L_*$ galaxies is about $4h^{-1}$Mpc.
For a Schechter function with a low-end slope of $-1$, one half of the galaxy mass is contributed by galaxies above $0.5L_*$.
Thus, if radio bubbles each have a mean radius of $0.5$Mpc for the galaxies considered, we have
$f=2\times 4\pi (0.5{\rm proper}\mpc)^3/3/(4h^{-1}(1+z)^{-3}\mpc)^3 \sim  14\%$ at $=2$, assuming that there is one occurrence of radio bubble event per $0.5L_*$ galaxy.
Adopting the peak value of $\eta_J$ shown in Figure (\ref{fig:Mhfjet}),
and assuming the average SMBH mass of the radio lobe launching AGNs is $10^8\msun$, corresponding to $L_*$ galaxies \citep[][]{2002Tremaine},
we obtain the number of pairs of radio lobes per galaxy $N_p=0.9 E_{60}^{-1}$,
where $E_{60}=E_R/10^{60}$erg with $E_R$ being the mean energy of each radio lobe;
the observed radio lobe total energy is in the range of a few times $10^{59}-10^{61}$erg \citep[e.g.,][]{2000Schoenmakers, 2001Kronberg}.
It thus seems that the overall volume filling fraction $f$ may fall in the range of order 10-20 percent.
In terms of galaxies, these regions around radio jet sources are likely highly biased.
But the fraction of
halos contained in these regions are likely substantially higher than $f$.
Therefore, we expect that the suppression effect due to magnetic pressure 
of accretion onto relevant halos will be substantial.
More importantly, the suppression effect will be spatially dependent,
potentially giving rise to a new kind of modulation of galaxy formation 
across different environments.

\subsection*{On the Global Downturn of Star Formation below $z=2-3$}

The observation that
this feedback effect strength due to magnetic pressure
becomes strong enough by $z=2-3$ to have crossed
a halo mass threshold of about $M_{\rm B}\sim 10^{12}-10^{12.5}\msun$ is significant.
This threshold halo mass is similar to the halo mass which separates cold and hot accretion modes \citep[][]{2005Keres}.
This implies that, while the halos more massive than about $M_{\rm B}\sim 10^{12.5}\msun$ will be
self-quenched due to lack of cooling, the smaller halos 
where cold accretion mode would otherwise
operate are now hindered from accreting gas due to magnetic pressure.
Consequently, star formation across the entire halo mass spectrum is now suppressed.
This may be a major culprit causing the
global downturn of star formation in the universe from $z=2-3$ to the present.

At this point, it is perhaps useful to clarify one point. 
This external, preventive feedback proposed here is not exclusive and does not rule out any possible internal feedback from AGN or from supernovae, both of which are undoubtedly present and important. 
In fact, in galaxies with sufficient cooling flows, internal AGN feedback has been demonstrated by many authors \cite{2004Omma, 2007Ciotti, 2015Choi} to be able to counter or delay cooling of the hot gaseous halo to significantly reduce star formation.

\subsection*{On the Flattening of Schechter Luminosity Function below $z=2-3$}

Another significant point to note is the rather prompt change in the observed 
slope of
the galaxy rest-frame UV luminosity function around $z=2-3$,
from $\alpha=-1.94$ at $z=2-3$
to $\alpha=-1.56$ at $z=1.0-1.6$ \citep[][]{2016Alavi}.
The traditional conjecture that supernovae
exert negative feedback on low mass galaxies would be in stark contradiction
to this observed change in galaxy luminosity function slope.
The argument goes as follows.
Star formation is the most vigorous at $z>2-3$ \citep[][]{1998Madau}.
Therefore, if supernova feedback were responsible for suppressing star formation 
in progressively lower mass galaxies,
the slope of the galaxy luminosity function at $z>2-3$ would have been
flatter than at lower redshift, when
star formation is less vigorous and hence suppressing effect on low-mass galaxies relatively less severe.
The negative effect due to magnetic pressure proposed here, however,
provides a natural explanation in timing.
Gas accretion suppression due to magnetic pressure becomes important only at $z<2$ or so
and the suppression is increasingly more severe on smaller mass halos, leading to a
flattening of the galaxy luminosity function at $z<2$.
As to how flattened the slope is due to this effect can not be easily estimated without detailed cosmological simulations.

\subsection*{An Argument For Preventive Feedback Processes}

Observations show a peak value of about 20\% of the global baryon to total mass ratio
around galaxies of halo mass $\sim 10^{12}$ in the SDSS galaxy sample \citep[e.g.,][]{2012Papastergis}.
This is tentalizing.
This is perhaps a direct piece of evidence against interval feedback as being the 
primary actor for driving gas away, simply because there is not an amount of
gas that corresponds to the global ratio to be driven away in the first place,
at least at relatively low redshift.
In other words, there is no widespread evidence of the existence of galaxies whose
baryon to total mass ratios are close to what the global ratio would indicate.
If our model bears the truth, one expects that such a "deficiency" of baryons 
should persist to $z\sim 2$ and then start to shift to more baryon rich galaxies at higher redshift, but the shift will be halo mass dependent with lower mass halos remaining
deficient longer.
Presently, most observational data at high redshift are presented 
as the gas to stellar mass ratio as a function of galaxy stellar mass
\citep[e.g.,][]{2017Genzel}.

\subsection*{Seeding Magnetic Field in Galaxies and Damped Lyman Alpha Systems}

If a significant volume of the intergalactic space can attain a magnetic field of a strength of $0.1\mu$G or so,
observed large magnetic field in some damped Lyman alpha systems at moderate redshift
may be accounted for;  for example, 
a simple contraction
from the mean intergalactic gas density of $\sim 10^{-5}cm^{-3}$ to a typical interstellar medium density of $1 cm^{-3}$ would amplify a $0.1\mu$G field to an amplitude of $200\mu$G, 
readily explaining an amplitude of $84\mu$G
in a galaxy at $z=0.7$ \citep[][]{2008Wolfe}.
Simulations have shown that damped Lyman alpha systems at moderate redshift of $z=2-3$
are largely caused by circumgalactic filaments and sheets
with a typical volumetric density in the range of $10^{-3}-10^{-1}cm^{-3}$
\citep[][]{2012Cen}.
Contraction alone from the mean intergalactic gas density of $\sim 10^{-5}cm^{-3}$
would result in a field in the range $2-50\mu$G for DLAs, consistent with the observed values of typically a few micro Gauss \citep[][]{1992Wolfe}.

\subsection*{Magnetic Field in the WHIM and Clusters}

In the preceding subsection, we discuss the overall magnetic field in the IGM that a Faraday rotation measure
based on line-of-sight integrals may yield.
An estimate of the magnetic field in filaments, i.e., the warm-hot intergalactic medium (WHIM)
at zero redshift \citep[][]{1999Cen}.
which make up a large portion of the missing baryons,
may be made.
Assuming that the magnetic field has largely been seeded due to the peak AGN activities at $z=2$,
then the magnetic field in filaments of density $n_e$ at $z=0$ is
\begin{equation}
\label{eq:BWHIM}
\begin{split}
B_{\rm WHIM}({\rm n_e}) = 0.66\mu{\rm G} \left({\bar B\over 0.1\mu~{\rm G}}\right) \left({n_e\over 10^{-4}\cm^{-3}}\right)^{2/3},
\end{split}
\end{equation}
assuming no further dynamo or other amplification.
With a $0.1-1\mu{\rm G}$ magnetic field expected in WHIM filaments,
if converging shocks compressing and forming filaments can accelerate or re-energize electrons,
filaments may be detected in synchrotron radiation by the next generation radio facilities, such as SKA.
Current observations have already begun to detect such emission from filaments even assuming magnetic fields weaker
than our estimates \citep[e.g.,][]{2021Vernstrom},
The expected field strength is
roughly consistent with synchrotron radiation observations, estimated based on energy equipartition assumption,
of the magnetic field in local filaments \citep[e.g.,][]{2019Govoni}.

In the cores of clusters of density $n_e\sim 10^{-3}-10^{-2}\cm^{-3}$ a field strength of several microGauss or tens of microGauss
is expected, even in the absence of any further amplification.
This is in line with observations \citep[e.g.,][]{2013Bonafede}.
What is perhaps more significant to note is that,
if the cluster center is anchored by a field strength of several microGauss or tens of microGauss,
a field strength of several $0.1\mu~{\rm G}$ to several microGauss may exist in the outskirts
of clusters of galaxies in our model, which appears to be observed \citep[e.g.,][]{2004Murgia,2013Bonafede}.

\subsection*{Cross-Correlations Between Faraday Rotation Measure and Others}

In our model, we expect that magnetic field injected into the IGM is most concentrated in proto-clusters.
As such, a significant cross-correlation signal is expected between proto-clusters and Faraday RM along a same line of sight.
To enhance the signal of this measure, one may denominate
the cross-correlation between proto-clusters and RM along the lines of sight through proto-cluster by
the cross-correlation between proto-clusters and RM along random lines of sight.
This may allow one to pick out the line-of-signt magnetic field strength in the proto-clusters fairly easily.

In addition to cross-correlating proto-clusters with Faraday RM to pick out
the magnetic field strength in the proto-clusters, one may also compute
the cross-correlation between Lyman alpha forest transmitted flux and RM.
The premise behind this method is that the Lyman forest transmission in proto-clusters of galaxies
is substantially different from the Lyman forest transmission in typical, average lines of sights.
This method shall yield RM measure as a function of Lyman alpha transmission flux.
As long as there is a difference in Lyman alpha transmission flux 
between proto-cluster regions and non-proto-cluster regions,
one may be able to teased out RM in proto-clusters in the future 
facilitated by SKA observations.

\subsection*{Deposition of Thermal Energy in IGM due to Expanding Lobe Shock Heating}

\begin{figure}
\centering
\includegraphics[width=1.0\linewidth]{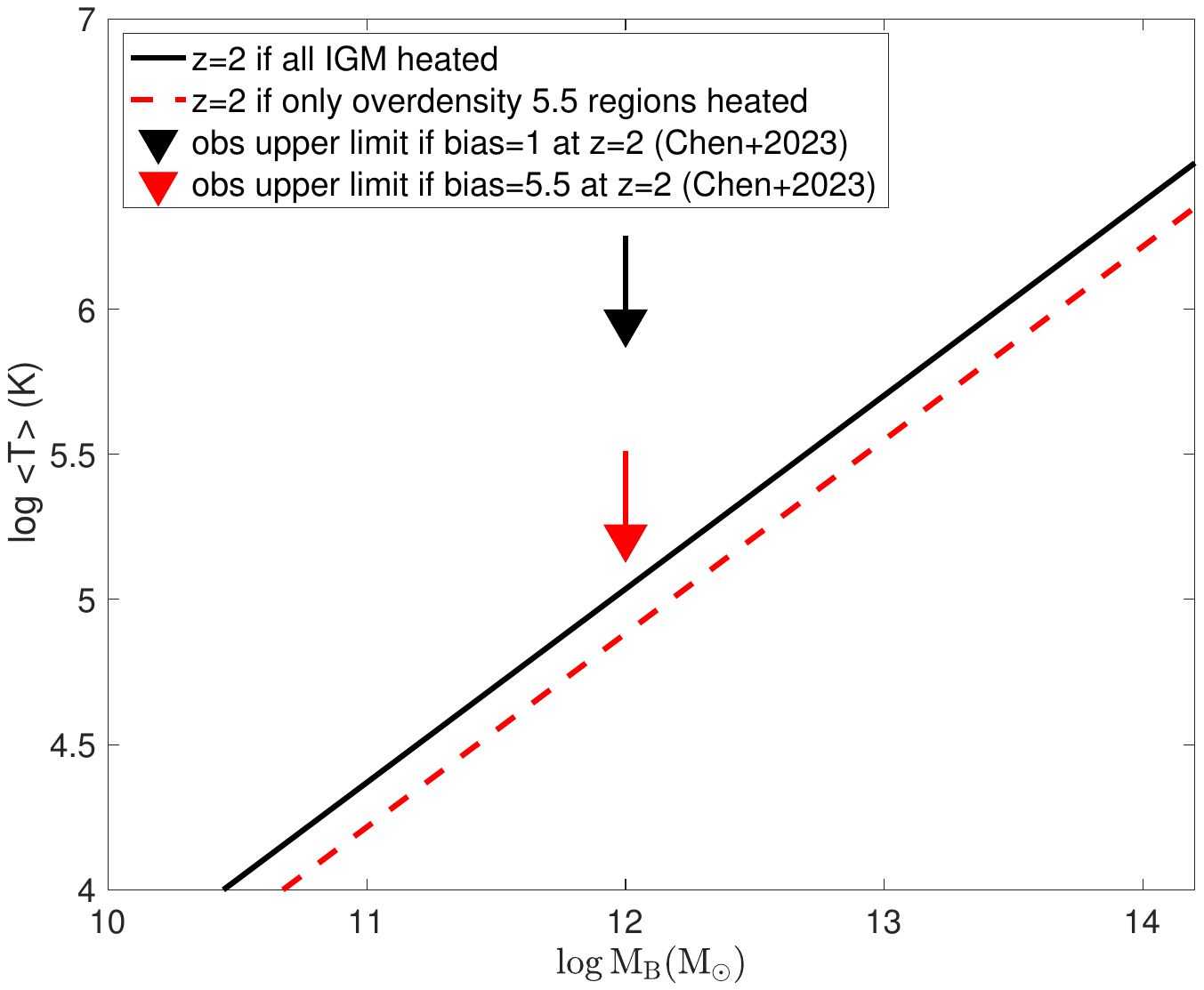}
\caption{shows the expected mean temperature per baryon in the entire universe as a function of
the magnetic Jeans mass $M_B$ at redshift $z=2$,
for two cases, one assuming that the entire IGM is heated (black solid line) and the other
assuming that only regions with overdensity of 5.5 is heated (red dashed line).
Also shown are upper limits observationally derived at $z=2$ 
\citep[][]{2023Chen} for two cases, 
one assuming that the entire IGM is heated with a bias factor equal to unity (blue arrow),
and the other assuming that only regions that have turned around from universal expansion
are heated with a bias factor equal to 5.5 (red arrow).
In shock propagation, zero metallicity for the ambient gas is assumed.}
\label{fig:TMB}
\end{figure}

The amount of thermal energy deposited in the IGM or proto-cluster regions or whereever
can be calculated.
We consider two regimes.
If the cooling time of the heated postshock regions due to the expanding lobe
is longer than the Hubble time at the redshift in question, then all thermalized energy
is counted.
On the other hand, if the cooling time of the heated postshock regions is shorter than
the Hubble time, the apparent amount of all thermalized energy will be reduced
by the ratio of the former to the latter, when one does a time averaging.
Assuming zero metallicity for the ambient gas into which the radio lobe driven shocks are propagating,
the results are shown in Figure (\ref{fig:TMB}),
where the amount of energy shown in the y axis is expressed as the mean temperature of the IGM
as a function of the magnetic Jeans mass $M_B$.
Two cases are shown, one assuming that the entire IGM is heated with a bias factor equal to unity (black solid line),
and the other assuming that only regions that have turned around from universal expansion
are heated with a bias factor equal to 5.5 (red dashed line).
In the latter case, the mean temperature is still averaged over all gas in the universe.
Also shown are upper limits observationally derived at $z=2$ 
\citep[][]{2023Chen}
for two cases, 
one assuming that the entire IGM is heated with a bias factor equal to unity (blue arrow),
the other assuming that only regions that have turned around from universal expansion
are heated with a bias factor equal to 5.5 (red arrow),
because the observationally derived upper limits depend on the bias factor of the heated region.
We see that, so long as $M_B\le 10^{12}-10^{12.5}\msun$,
the thermal energy injected by expanding giant radio lobes 
is consistent with current observational limit.

\subsection*{Magnetic Field At Virial Radius as a function of Halo Mass}

For halos less massive than the magnetic Jeans mass $M_B$,
we expect that accreting gas will be prevented from entering the virial radius.
As a result, the field may accumulate outside the virial radius of such a halo.
Consequently, we expect that the magnetic field
strength at the virial radius (assuming overdensity of 100)
will be just at the threshold level for each halo at lower redshift at
\begin{equation}
\label{eq:Bh}
\begin{split}
B_{\rm h}(\sigma_v) = \sqrt{600\pi \Omega_B\rho_{c,0} (1+z)^3} \sigma_v,
\end{split}
\end{equation}
where $\sigma_v$ is the 1-d velocity dispersion of the halo, $\rho_c$ is the critical density
of the universe at $z=0$, and $\Omega_B$ baryon density.
This prediction is potentially testable by SKA in the future.

\subsection*{Suggestion on How to Implement This MHD Effect}

A simple version of how to implement this MHD effect may be described as follows, although more sophisticated implementation may be devised.  
First, one runs radio jet expansion simulation not in a cosmological simulation box but in an isolated box before hand to produce a standard mold of magnetic field structures in a bubble of size 1 proper Mpc, to be inserted into the cosmological simulation. One may inject a prescribed amount of magnetic energy into the center over 100Myr to mimic FR II sources. 
Second, one runs a cosmological MHD simulation of a sufficiently large box to contain one cluster of galaxies by $z=0$. One then retraces back to the starting redshift to identify a region that contains the proto-cluster of the cluster found at $z=0$. Then, one re-runs the simulation with static meshrefinement in the zoom-in region that contain the proto-cluster from the starting redshift. At $z=2$, one adds a pair of radio bubbles of radius 1 proper Mpc for each of the masssive galaxies at a distance of 1Mpc from that galaxy in the proto-cluster using the mold in the first step with appropriate adjustments in velocity and density field based on the local values in the cosmological simulation box, and then re-starts the simulation from $z=2$ to run to $z=0$.

\section*{Conclusions}

Utilizing a set of parameters fully anchored on observational data, 
we show that giant radio lobes from supermassive black holes residing in massive galaxies
can inject enough magnetic energy so as to exert a major negative feedback effect,
by significantly retarding or preventing gas accretion onto halos as massive as 
$M_{\rm B}=10^{13}\msun$ by $z=2$, depending on the volume filling fraction of the radio bubbles
in the universe, thanks to the magnetic pressure of the accreting gas. The accretion suppression effect increases with decreasing halo mass. 
We shall call $M_B$ the magnetic Jeans mass.

The implication is profound.
That is, by $z=2-3$, halos where cold accretion mode would otherwise
operate are now significantly prevented from accreting gas due to magnetic pressure.
This negative feedback process may be the culprit for 
causing the global downturn in star formation 
in the universe from $z=2-3$ to the present.
This feedback mechanism also provides a natural explanation 
for the rather prompt change in the slope of
the galaxy rest-frame UV luminosity function around $z=2-3$,
from $\alpha=-1.94$ at $z=2.2-3$
to $\alpha=-1.56$ at $z=1.0-1.6$ \citep[][]{2016Alavi}.

Finally, a number of ramifications and predictions due to this process
are given.
First, magnetic fields for a host of systems,
such as galaxies and damped Lyman alpha systems at moderate redshift,
and
extragalactic filaments and clusters of galaxies low redshift,
may be accounted for.
Second, the thermal energy injected by shocks due to supersonically expanding 
magnetic bubbles is significant but in agreement with current observational
upper limits.
Third, cross-correlations between proto-clusters and Faraday rotation measures
should be able to test the predicted magnetic field directly. 
Cross-correlations between Faraday rotation measure and Lyman alpha forest flux
spectrum can provide additional information on this.

\matmethods{
A combination of empirical observational data and bubble evolution analytics 
is used to quantify a potentially significant preventive effect on 
the dynamics of intergalactic gas accretion onto dark matter halos.
We make use of the observed abundance of radio galaxies, the observed fractional energy
released in the form of large radio jets in terms of supermassive black rest mass energy
and the total amount of mass in supermassive black holes in units of the stellar mass formed in the universe.
We deduce the total amount of magnetic energy in giant radio lobes based on these observational data,
in combination with results obtained based on MHD radio bubble expansion simulations that estimate
the surviving magnetic energy when realistic bubbles reach a radius of order 1Mpc.
Combining these results, we estimate the magnetic pressure of gas accreting into dark matter halos
and show that it could provide a significant hinderance for gas accretion, thus potentially
preventing or slowing or retarding gas accretion onto dark matter halos as massive as $10^{12}\msun$
at $z<2-3$.
Uncertainties and spatial variations of this effect is discussed, depending the volume filling factor
of the radio bubbles.
}
\showmatmethods{} 

\acknow{
This work is supported in part by
the National Key Research and Development Program of China.
The author thanks Princeton University for hospitality during a visit when this work was completed.
I thank Romain Teyssier for discussion.
}



\bibsplit[2]


\end{document}